# Sub-$10 sound card photogate variants

Zoltan Gingl
Department of Experimental Physics, University of Szeged
Dóm tér 9. Szeged, H-6720, Hungary

Recently, a simple and very low-cost photogate has been shown by Horton[1] as an efficient experimentation tool in physics education. The photogate connects to the microphone input of a personal computer and a free software can be used to visualize the light interruptions caused by a moving object like a pendulum. Although the device works properly, there are further possibilities of improvement and similar alternatives also exist. The following brief review may help teachers to pick the one that best fits their needs and possibilities.

## The simplest photogates ever

The photogate described by Horton is based on a photoresistor[2], whose resistance strongly depends on light intensity. Since the microphone input can measure voltage only, a battery-driven voltage divider configuration was used to provide voltage depending on the light intensity. On the other hand, PC microphones need bias voltage, therefore the microphone connector of the sound card has a terminal coupled to a voltage source via a series resistor. The connections are not the same for all computers: Figure 1 illustrates the possibilities.

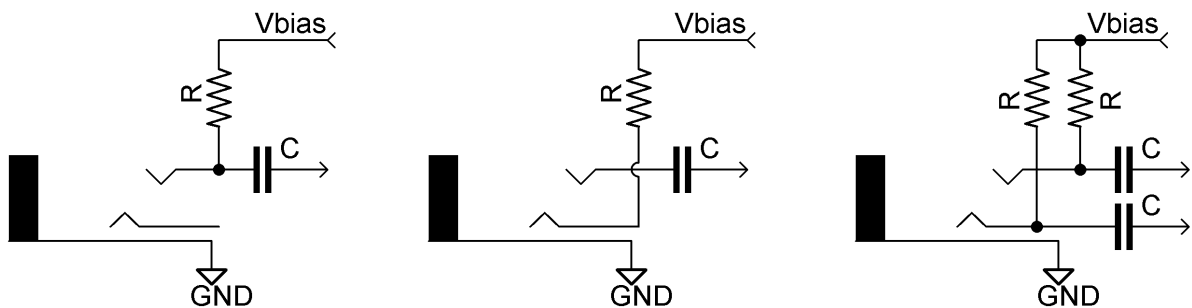

**Fig.1. Simplified microphone input connection possibilities. Typical values are: Vbias=2.5V to 5V, R=2k, C=1uF. The left two configurations are the most frequent; the "true stereo" microphone input is shown on the right.**

This means that the external battery and resistor are not needed: we can get an extremely simple photogate by just connecting the photoresistor directly to the microphone input, as show on Figure 2. The computer's internal bias voltage and series resistor form a light-controlled voltage divider with the externally connected photoresistor. This connection will work for all microphone input configurations. Thanks to the minimal number of components, it is very easy to make, very reliable and costs no more than $2.

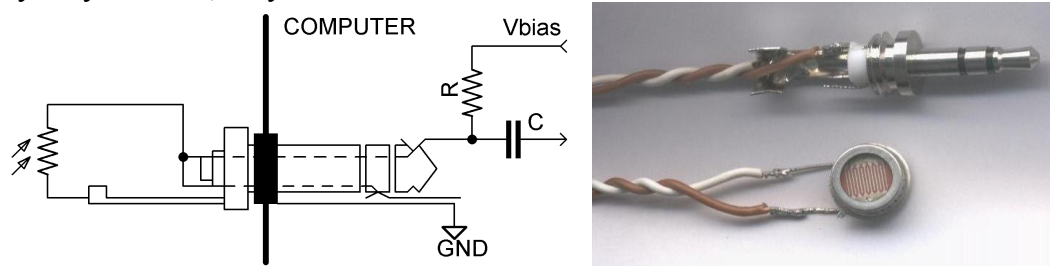

**Fig.2. The photoresistor (WK65060), the microphone input's internal bias voltage (typically 2.5V) and the series R=2k resistor form a resistor divider. The tip and ring of the 3.5mm jack plug and one pin of the photoresistor should be tied together, as the sleeve and the other pin of the photoresistor.**

Photoresistors are not the only sensors that can be used to detect light. Professional photogates employ phototransistors[3]; they conduct better in the presence of light. Phototransistors outperform photoresistors in several aspects. They switch much faster (within a few microseconds, while photoresistors have rise and fall times in the range of 25ms to 100ms), their sensing surface is smaller (the spatial resolution is better), they are broadly available in different wavelengths and sensitivities, they are offered in optically filtering packages, well documented, cheap and easy to purchase. Figure 3 shows that this sensor can also be connected directly to the microphone input; the phototransistor works in a common emitter configuration. Phototransistors may have a price down to half a dollar, so the overall cost can be kept below $2 again.

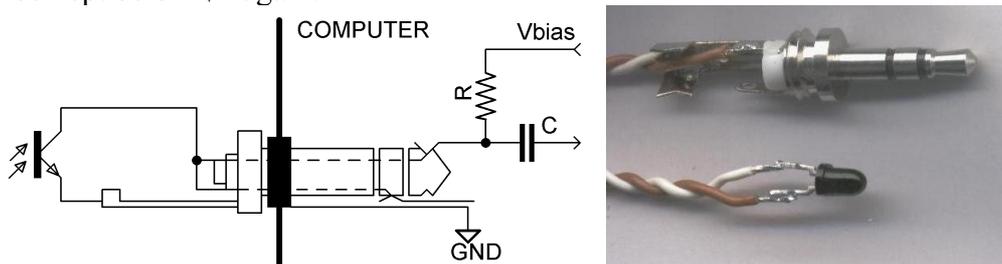

**Fig.3. The phototransistor can be connected directly to the microphone input. The tip and ring of the 3.5mm jack plug are connected to the collector, the sleeve is soldered to the emitter. The SFH309F infrared phototransistor looks black, due to its optically filtering package.**

If for any reason, one can't connect the photoresistor or phototransistor to a microphone input, it is still possible to make a resistane-to-voltage conversion without the need of external battery. The USB port has a 5V supply line, so the risk of discharged batteries, contact problems can be reduced as shown on Figure 4. Although the USB power can be somewhat noisy, it is still adequate to bias the photogate.

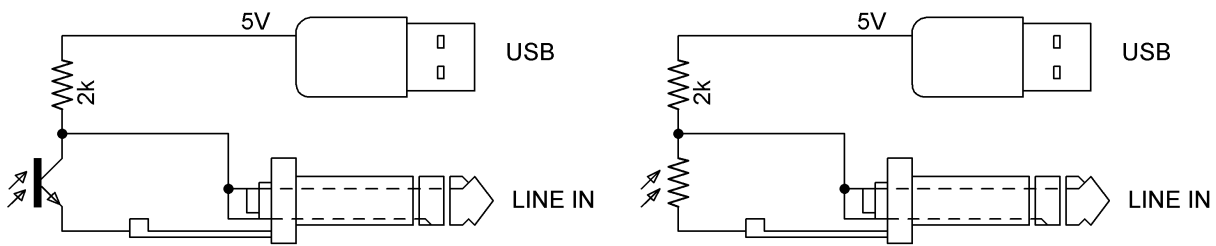

**Fig.4. The phototransistor and photoresistor can be connected to the line input, if the USB port provides the required bias voltage.**

## Experimental demonstration

The performance of the photogates based on phototransistors and photoresistors directly connected to the microphone input has been tested and compared by detecting the free fall of a picket fence often used to determine acceleration due to gravity.[4] A flashlight served as light source, and the output signals of the photogates were recorded. The results plotted on Figure 5 demonstrate the highly accurate response of the phototransistor. They also suggest that the phototransistor is more sensitive. Using the phototransistor the instantaneous velocity[5] can be precisely determined.

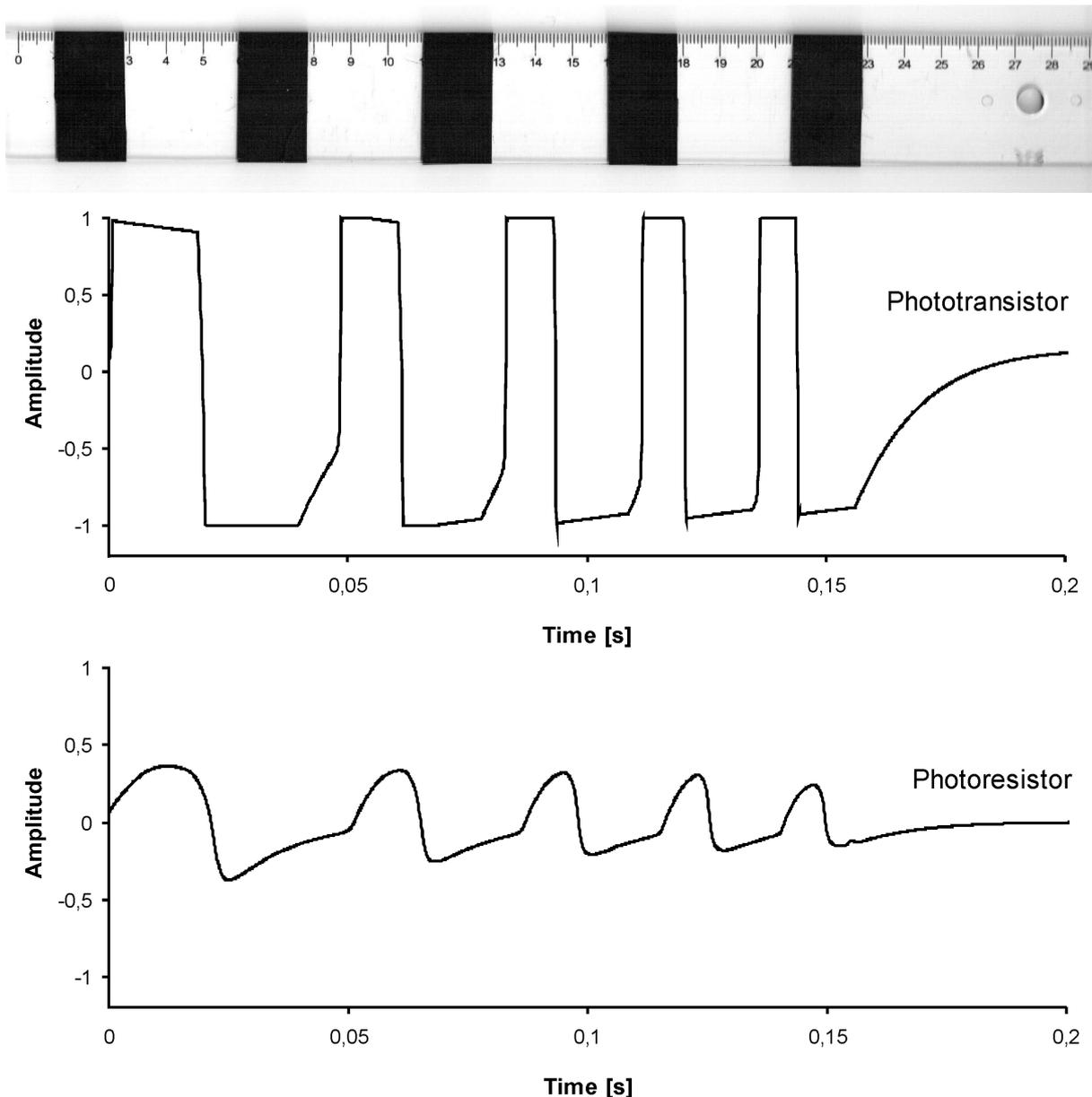

Fig.5. Photogates detected the free fall of a picket fence made using a liner and insulating tape. A flashlight served as light source, a TEPT4400 ambient light phototransistor and a WK65060 photoresistor were used as light detectors. Both sensors were directly connected to the microphone input of the computer and the same microphone volume setting has been used.

## Conclusions

Very simple photogates can be made easily by connecting a single phototransistor or photoresitor directly to the microphone input of a personal computer – no additional components are needed. Although photoresistors work well in most cases, superior performance and better availability make phototransistors the best choice. The cost of such photogates is incredibly low, can be less than $2. Thanks to their simplicity, these photogates are very reliable and easy to use. Students can make their own photogates without considerable effort, can do the experiments at home as a part of a homework or just can have fun by playing with their computer as a motion or light detector. These result in better motivation, more room for creativity.

Finally, an important warning for all experimenters: it is a must to isolate all electrical connections from the user with insulating tapes or heat shrink tubes. User's safety and accidental damage to the electronic components via electrostatic discharge (ESD) require the protection from directly touching the wires.

Additional information can be found at http://www.noise.physx.u-szeged.hu/edudev/.